\documentclass[twoside]{ilcws10}
\usepackage[latin1]{inputenc}
\usepackage{graphicx}
\usepackage{wrapfig,rotating}
\usepackage{amssymb,amsmath,array}

\begin{document}
\title{DEPFET Vertex Detectors: Status and Plans}
\author{Frank Simon$^{1,2}$, for the DEPFET Collaboration
\vspace{.3cm}\\
1- Max-Planck-Institut f\"ur Physik, Munich, Germany
\vspace{.1cm}\\
2- Excellence Cluster `Universe', TU M\"unchen, Garching, Germany
}

\maketitle

\begin{abstract}
DEPFET active pixel sensors are a well-developed technology for vertex detectors at future colliders. Extensive test beam campaigns have proven the excellent performance of these devices, and their radiation hardness has been thoroughly tested. For the Belle-II experiment at the SuperKEKB collider, a new vertex detector based on DEPFET technology is being developed, using sensors thinned down to 75 $\mu$m. We give an overview over recent results with test devices using ILC pixel geometries as well as the concepts and challenges for the Belle-II pixel vertex tracker and discuss how the R\&D for the ILC VXD can take advantage of these developments. \end{abstract}

\section{Introduction: DEPFET Pixel Detectors for ILC and Belle-II}

Vertex detectors are crucial components of modern particle physics experiments. They are indispensable for the identification of heavy quarks, which is based on the reconstruction of a separate decay vertex for longer-lived hadrons containing these quarks. At a future linear collider, the correct identification of charm and bottom quarks is particularly important for precision Higgs and Top physics as well as for physics beyond the Standard Model. To achieve the goals of the impact parameter resolution for secondary tracks set by these physics requirements, a highly efficient low-mass vertex detector with a single hit resolution exceeding 5 $\mu$m is needed. 

\begin{wrapfigure}{r}{0.4\textwidth}
\centering
\includegraphics[width=0.38\textwidth]{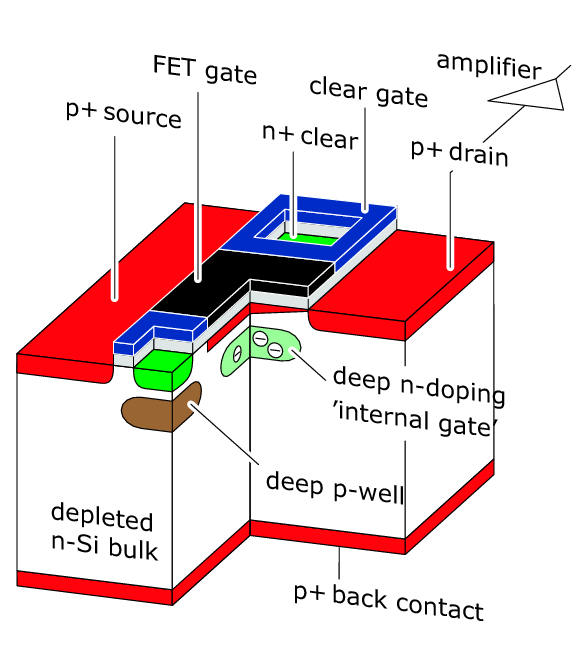}
\caption{Illustration of the layout of one DEPFET pixel.}\label{fig:DEPFET}
\end{wrapfigure}

One of the technologies able to match these requirements is the DEPleted Field Effect Transistor (DEPFET) \cite{Kemmer:1986vh}, an active pixels sensor with p-channel FETs integrated in a fully depleted bulk, illustrated in Figure \ref{fig:DEPFET}. The charge of a through-going particle is collected in an internal gate created by an n-inplant beneath the transistor channel, leading to a modulation of the channel current of the transistor depending on the amount of collected charge. These devices thus provide in-pixel amplification, combined with low noise due to the small capacitance of the internal gate. The amplification combined with the low noise leads to excellent signal to noise ratios, allowing the construction of thinned detectors to reduce the material budget. 

The power consumption of the active area of DEPFET sensors is very low since the pixels passively collect charge and only need power during the readout cycle. The pixels are read out row-wise in a rolling shutter mode and are cleared after readout. 

Recently, the DEPFET technology has been chosen for the pixel vertex detector of the Belle-II experiment \cite{Adachi:2008da, Dolezal:2009wq}, which will be located at the future SuperKEKB collider \cite{Hashimoto:2004sm, Dolezal:2009wq}, an upgrade of the asymmetric energy KEKB flavor factory. This super flavor factory will provide luminosities of up to $8\,\times\,10^{35}\,\text{cm}^{-2}\text{s}^{-1}$ and will run predominantly at the  $\Upsilon(4s)$ resonance at a center of mass energy of 10.58 GeV. The vertexing at such a machine is crucial for the measurement of the decay time difference between two $B$ mesons, obtained from the spatial separation of the two decay vertices. In the case of $B^0$ pairs, the decay time differences are typically of the order of 1.5 ps, corresponding to flight distances of around 200 $\mu$m along the beam axis.

The requirements for a vertex detector at Belle-II and at a linear collider differ in some aspects, due to the differences in typical particle momentum, which is much lower at Belle-II than at the ILC, and due to differences in the background and collision rates. Here, the quasi-continuous beam with 2 ns between bunch crossings and 100\% duty cycle combined with high background rates at Belle-II requires fast continuous readout. At the ILC, the more than two orders of magnitude larger bunch to bunch spacing, combined with the bunch train structure of the beam, place less stringent requirements on the readout speed of the detector. In both cases, the need for a low material budget to limit multiple scattering precludes active cooling beyond flowing of cold gas in the active area of the detector, placing limits on the overall power consumption of the sensors. At ILC, a power pulsing of the electronics is possible, while continuous running is required at Belle-II.  

These requirements allow for less pixels per module in the case for Belle-II compared to the ILC case, leading to different pixel sizes in the detector design. To maximize the spatial resolution, a pixel size of around  $25\,\times\,25\,\mu\text{m}^2$ is envisioned for the vertex detector at a linear collider, while the Belle-II design will use pixel sizes of  $56\,\times\,50\,\mu\text{m}^2$ and $77\,\times\,50\,\mu\text{m}^2$. To minimize the material budget, the ILC sensors will be thinned down to 50 $\mu$m in the active area, while for Belle-II a thickness of 75 $\mu$m is being considered, which optimizes the spatial resolution achieved through charge sharing between pixels while keeping the contributions of multiple scattering in the detectors negligible compared to scattering in the beam pipe.  

In the overall system including sensors and on- and off-detector electronics, many similarities between a DEPFET pixel detector for the ILC and the Belle-II vertex detector exist. The construction of the Belle-II detector thus leads to an accelerated development which also benefits the studies for the ILC.

\section{Test Beam Results}

Since 2005, various different DEPFET detector prototypes have been tested in a 2\,--\,6 GeV electron beam at DESY, in a 24 GeV proton beam at the CERN PS and in a 40\,--\,120 GeV pion beam at the CERN SPS \cite{Velthuis:2008zza,Moser:2007zz,Andricek:2009zz,Marinas:2009su}. For these tests, several devices were operated as a telescope, with the central detectors used as devices under tests, mounted on precision translation and rotation stages. In the most recent test beam campaigns, unthinned sensors with 450 $\mu$m thick silicon were tested. For these tests, devices with $64\,\times\,256$ and $64\,\times\,128$ pixels with ILC-type pixel sizes, ranging from $20\,\times\,20\,\mu\text{m}^2$ to $24\,\times\,32\,\mu\text{m}^2$, were used. 

With the track information provided by the beam telescope made up out of the other modules in the test, which achieves sub-micron precision, the resolution of a DEPFET sensor with $24\,\times\,24\,\mu\text{m}^2$ pixels was found to be
\begin{equation*}
\sigma_x\, = \, 1.3\, \mu\text{m}\, \pm \, 0.1 \,\mu\text{m}, \hspace{5mm} \sigma_y\, = \, 1.2\, \mu\text{m}\, \pm \, 0.1 \,\mu\text{m},
\end{equation*}
demonstrating the excellent performance of these devices for high-precision vertex detectors.  

Signal to noise ratios of up to 200 for 450 $\mu$m thick sensors with DEPFET pixels  with a capacitatively coupled clear gate with a length of 5 $\mu$m were achieved, showing a comfortable margin also for thinned sensors with correspondingly smaller signal charge but constant noise levels. 

\begin{wrapfigure}{r}{0.5\textwidth}
\centering
\includegraphics[width=0.44\textwidth]{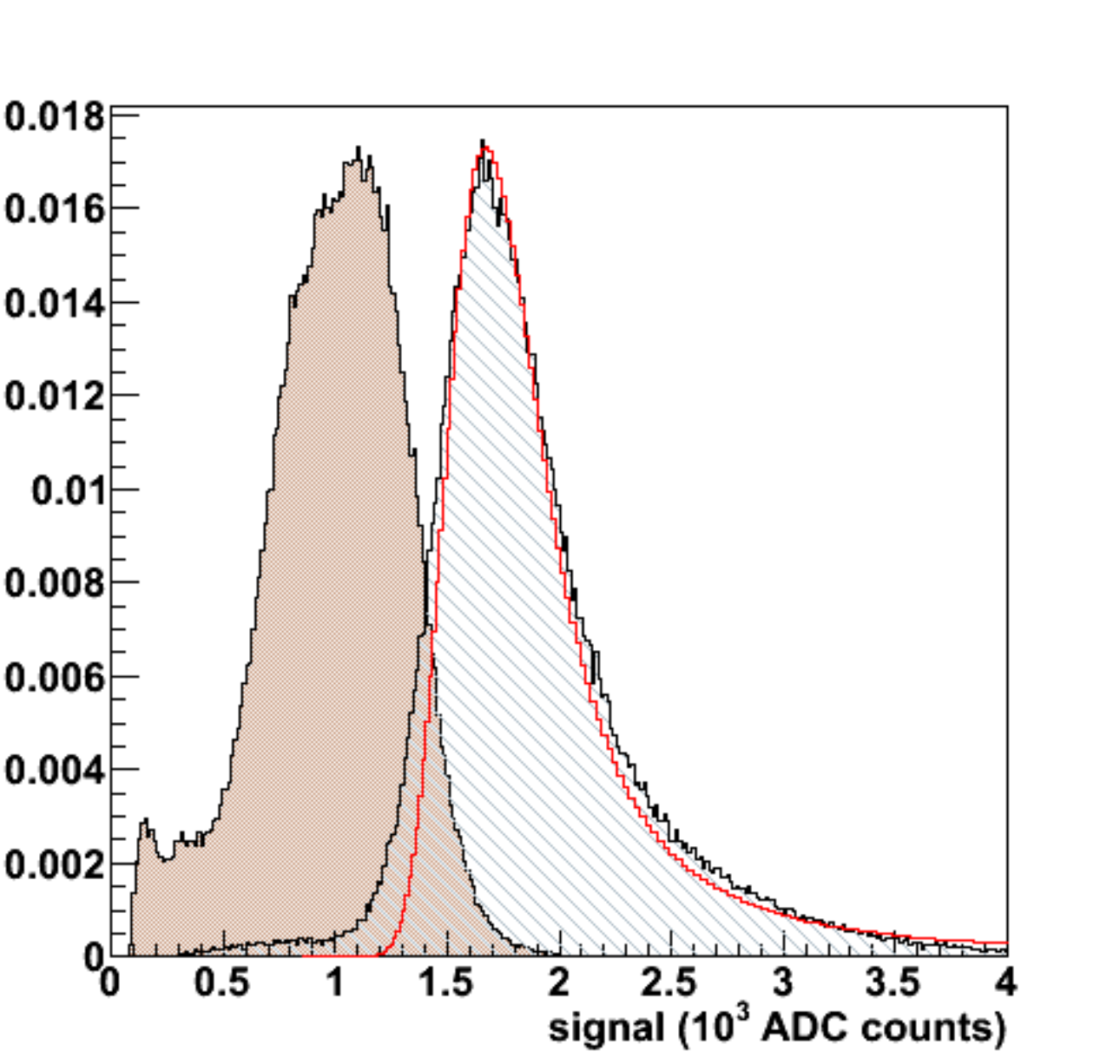}
\caption{Distribution of cluster(light gray) and seed pixel (brown) amplitudes. The cluster amplitude follows the expected Landau distribution (red).}\label{fig:Cluster}
\end{wrapfigure}

The wealth of data available from the test beams was also used for detailed studies of the signal of the detectors.  Figure \ref{fig:Cluster} shows the distribution of the amplitude of reconstructed clusters and of the amplitude of the corresponding seed pixel. The clustering algorithm uses a seed threshold of 7\,$\sigma$ and includes pixels above the neighbor threshold of 2.6\,$\sigma$ in the cluster. The cluster amplitude follows the expected Landau distribution for charge deposits in thin layers. 

The test beam data also provides important input for the development and for the validation of detector simulations. These simulations are crucial to evaluate the performance of complete pixel vertex trackers in the design phase, and are used both for design studies for a vertex tracker at a future Linear Collider experiment and for the optimization of the detector parameters for the Belle-II experiment. 

\begin{figure}[!hb]
\centering
\includegraphics[width=0.85\textwidth]{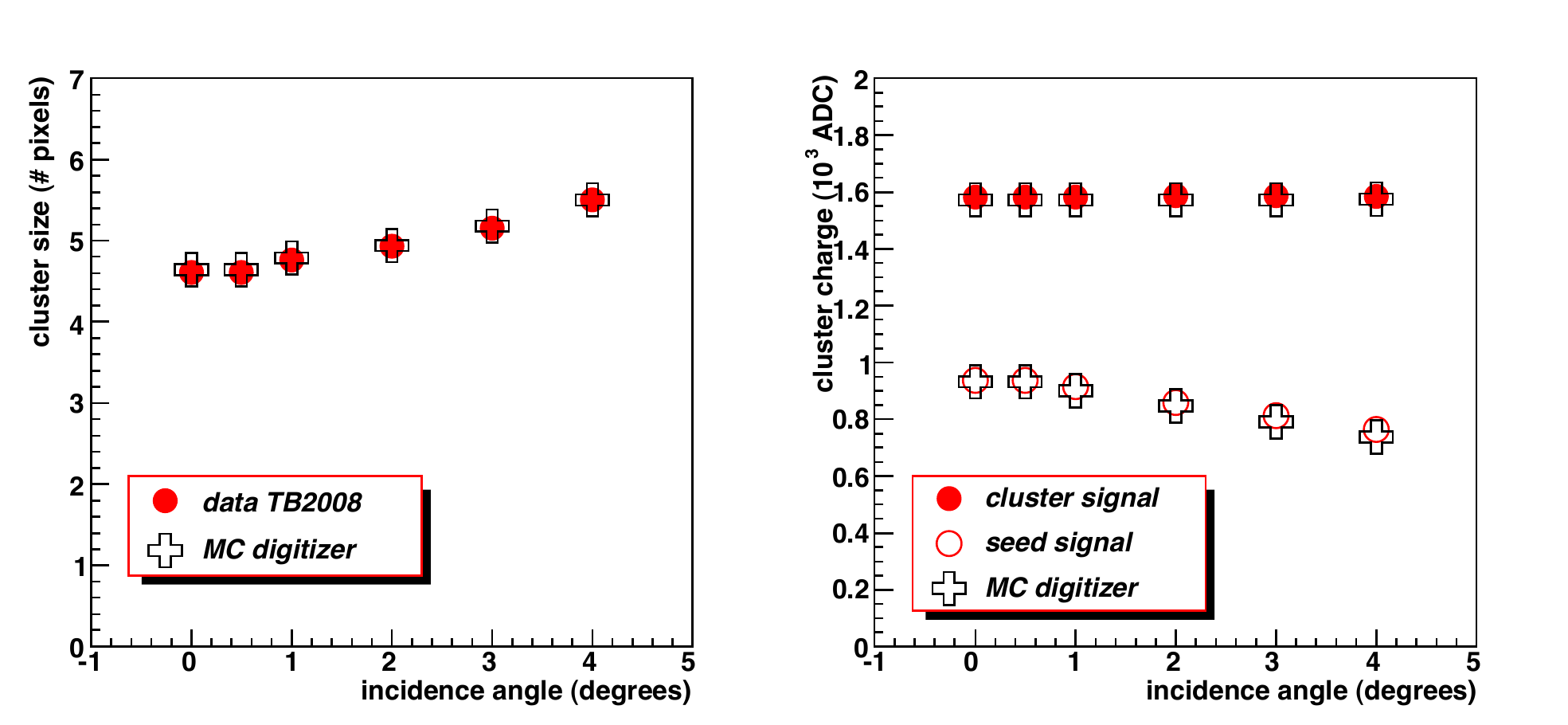}
\caption{Cluster size (left) and cluster and seed pixel charge (right) as a function of the track incidence angle, shown for data and for Geant4 simulations including detector digitization.}\label{fig:ClusterDigitizer}
\end{figure}

Figure \ref{fig:ClusterDigitizer} shows measurements of the cluster size in pixels and the cluster and seed pixel charge as a function of incidence angles for small variations around normal incidence. These results are compared to simulations based on Geant4 with digitization, a full description of the detector response. The excellent agreement demonstrates the maturity and reliability of the simulations used for the optimization of DEPFET-based detector systems.

\section{The Pixel Vertex Detector for Belle-II}

\begin{figure}
\includegraphics[width=0.9\textwidth]{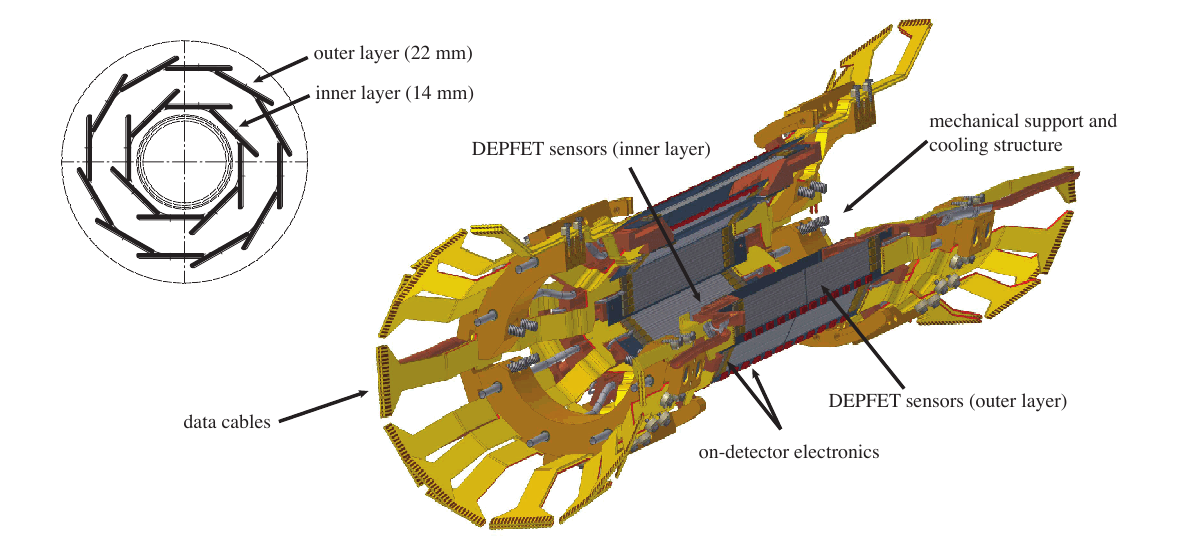}
\caption{Illustration of the Belle-II PXD, showing the arrangement of the modules of the inner and outer layers (inset upper left), and a the schematic of the overall mechanical design, including support and cooling structures and data cables.}\label{fig:PXD}
\end{figure}

The new pixel vertex detector (PXD) for Belle-II will consist of two layers, at a radius of 14 mm and 22 mm, respectively, arranged around the thin straight section of the Beryllium beam pipe in the interaction region of the Belle-II detector, as illustrated in the inset in the upper left corner of Figure \ref{fig:PXD}. The inner layer uses 8, the outer layer 12 individual detector modules, with an active length of 90 mm for the inner and 123 mm for the outer layer, sufficient to cover the full acceptance of the Belle-II detector from 17$^\circ$ (forward, in the direction of the high energy beam) to 150$^\circ$, relative to the axis of the solenoidal field in the Belle-II detector. 

\begin{wrapfigure}{r}{0.5\textwidth}
\includegraphics[width=0.4\textwidth]{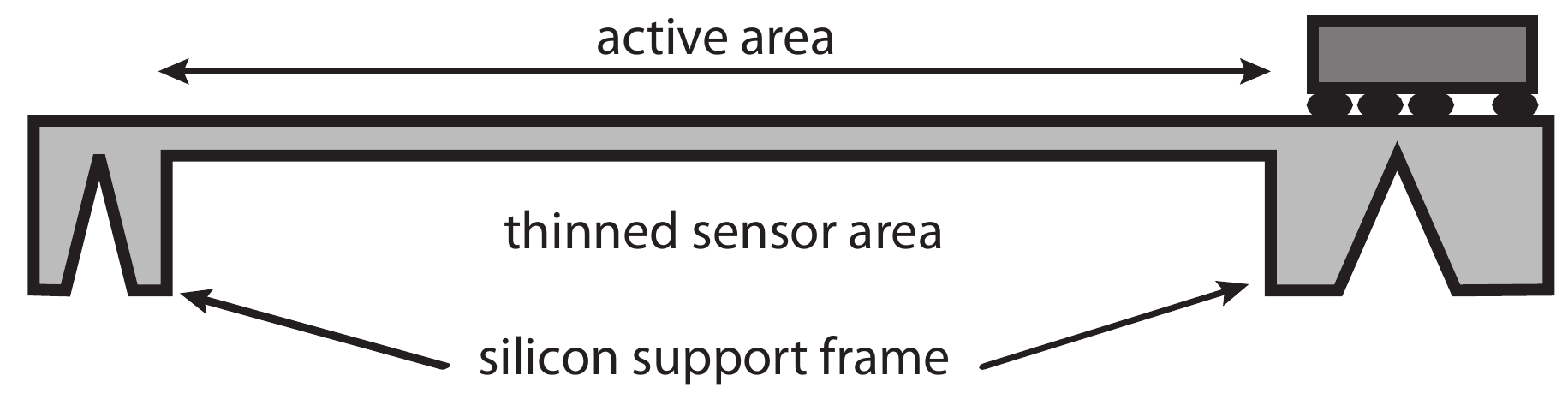}
\caption{Illustrative cross section through a PXD module. The silicon in the active area is thinned to 75 $\mu$m, 450 $\mu$m thick silicon rims are left on the module sides for stability.}\label{fig:CrossSection}
\end{wrapfigure}

To maximize production yield and to satisfy the space constraints from the 6'' wafers used or the sensor production, the modules for both the inner and the outer layer will be constructed from two sensors, which will be joined in the middle. To minimize the amount of material, crucial for the precise reconstruction of low momentum tracks, the modules themselves are thinned down to a thickness of 75 $\mu$m in the active area, with 450 $\mu$m thick silicon rims to provide structural stability, as illustrated in Figure \ref{fig:CrossSection}. Thus, no additional mechanical support within the detector acceptance is necessary.

Both inner and outer layer sensors will have 1600 by 250 pixels, which are continuously read out in a rolling shutter mode, with a readout time of 20 $\mu$s. Given the shorter length of the modules of the inner layer, smaller pixels, with a size of $56\,\times\, 50 \, \mu\text{m}^2$ are used close to the interaction point, while the outer modules have slightly elongated pixels with a size of  $77\,\times\, 50 \, \mu\text{m}^2$. The width of the sensitive area of both the inner and the outer modules is 12.5 mm and the total width, including support rims, is 15 mm. The read out electronics are located on the module ends, with additional chips for the addressing of individual pixel rows mounted on one side of the stave. The total power consumption is around 7 W per module side, with about 0.5 W dissipated over the active area, and 0.5 W dissipated in the switcher electronics on the stave edges. The required cooling capacity imposes strict constraints on the detector support. The still evolving design is based on support rings on both detector ends made of copper, chosen for the excellent thermal conductivity, which hold the silicon modules. The rings themselves contain cooling channels for active cooling, and through-going air channels to provide cooling for the sensor surfaces with a cold air flow. Several options for the cooling of the support rings are under investigation, with evaporative CO$_2$ cooling as the preferred technology because of the large cooling power combined with minimal space requirements for cooling pipes and heat exchangers in the mechanical support. Alternatives include water cooling and evaporative cooling using C$_3$F$_8$. The current mechanical design of the PXD, including support structures and data cables, is illustrated in Figure \ref{fig:PXD}.

Due to the high number of pixels and the continuous readout in a background dominated environment, the zero suppressed data rate of the PXD is expected to be on the order of 10 GB/s, far beyond the data rate of all other detector systems in Belle-II combined. A data reduction scheme using fast online tracking in the silicon strip detector to select regions of interest in the PXD for further data processing has been developed to reduce the PXD data output to a manageable level. First simulation studies of such a scheme show promising results, and will be pursued further.

The final detector parameters are being determined with detailed single track and physics simulations. First mechanical prototypes have been constructed to prove key concepts of the mechanical design, such as the low mass connections of two sensors into one module or the fixation of the silicon to the mechanical support. Further studies of the mechanics and of the cooling structures are in preparation. The assembly and installation of the PXD in the Belle-II detector and the beginning of data taking at the SuperKEKB flavor factory is envisaged for the year 2014.

\section{Summary and Outlook}

The DEPFET is a mature technology for high-resolution, low-mass pixel vertex detectors. It has been thoroughly studied in several test beam campaigns, demonstrating an excellent signal to noise ratio and very high spatial resolution. This technology was chosen for the new vertex detector of the Belle-II experiment, and is able to fulfill the requirements for vertexing at the ILC. A test production of thinned DEPFET sensors with the Belle-II PXD and ILC pixel geometries is well advanced, and will undergo thorough testing in the near future to demonstrate the performance of 50 $\mu$m thick sensors. Beam tests with these new devices are planned for fall 2010 at CERN. With the construction of a full detector based on DEPFET technology, the development of all components, ranging from sensors to on- and off-detector electronics, data acquisition and mechanical support, is significantly accelerated, benefitting also the design of an ILC vertex detector based on this technology.


\begin{footnotesize}


\end{footnotesize}


\end{document}